\documentclass[preprint,showpacs,preprintnumbers,amsmath,amssymb]{revtex4}

\usepackage{graphicx}
\usepackage{dcolumn}
\usepackage{bm}

\begin{document}


\title {Possible effects of tilt order on phase transitions of a fixed connectivity surface model}

\author{Hiroshi Koibuchi}
 \email{koibuchi@mech.ibaraki-ct.ac.jp}
\affiliation{%
Department of Mechanical and Systems Engineering, Ibaraki National College of Technology, Nakane 866 Hitachinaka, Ibaraki 312-8508, Japan}%

\date{\today}

\begin{abstract}
We study the phase structure of a phantom tethered surface model shedding light on the internal degrees of freedom (IDOF), which correspond to the three-dimensional rod like structure of the lipid molecules. The so-called tilt order is assumed as IDOF on the surface model. The model is defined by combining the conventional spherical surface model and the $XY$ model, which describes not only the interaction between lipids but also the interaction between the lipids and the surface. The interaction strength between IDOF and the surface varies depending on the interaction strength between the variables of IDOF. We know that the model without IDOF undergoes a first-order transition of surface fluctuations and a first-order collapsing transition. We observe in this paper that the order of the surface fluctuation transition changes from first-order to second-order and to higher-order with increasing strength of the interaction between IDOF variables. On the contrary, the order of collapsing transition remains first-order and is not influenced by the presence of IDOF.

\end{abstract}

\maketitle
\section{Introduction}\label{intro}
The crumpling transition has long been interested in membrane physics and in biological physics \cite{NELSON-SMMS2004,Gompper-Schick-PTC-1994,Bowick-PREP2001}. The curvature model of Helfrich, Polyakov and Kleinert \cite{HELFRICH-1973,POLYAKOV-NPB1986,KLEINERT-PLB1986} for membranes was found to undergo  first-order transitions on spherical and fixed connectivity surfaces by Monte Carlo (MC) simulations \cite{KD-PRE2002,KOIB-PRE-2004-1,KOIB-PRE-2005,KOIB-NPB-2006}.

Internal degree of freedom (IDOF) corresponding to the three-dimensional rod like structure and  electrostatic structures such as a dipole moment of molecules are out of consideration in the curvature models. The surface models are those defined only by two-dimensional differential geometric notions \cite{DAVID-SMMS2004}. For this reason, the thermodynamic properties of the models can easily be accessed so far in theoretical/numerical studies \cite{Peliti-Leibler-PRL1985,DavidGuitter-EPL1988,PKN-PRL1988,KANTOR-NELSON-PRA1987,AMBJORN-NPB1993}. 

However, three-dimensional structure of molecules is considered to play important roles in specific phenomena in membranes. On the Langmuir monolayer, photoinduced traveling waves were observed experimentally \cite{TYY-NJP-2003}. The traveling wave is carried by the rotation of the molecular azimuth, where the tilt angle is kept constant. This clearly indicates that the molecules tend to align to each other. The chirality of membranes is also considered to be connected to the tilt of lipids. The directional order-disorder transition corresponding to such three-dimensional structure of molecules is the so-called gel-liquid crystal transition, which can be observed in bilayers including biological membranes. Lippling transition \cite{LL-PRB1987} is also considered to be connected to the IDOF such as directional order-disorder of the lipid molecules. Moreover, it is also quite well known that a variety of shapes and topology of membranes are both closely related to internal molecular structures. In facts, lamellar, hexagonal, and vesicles are understood to be originated from the difference in the shape of lipids.

The tilted molecules and its relation to the shape and the chirality of membranes have long been studied \cite{HelfrichProst-PRA1988,ZJX-PRL1990,SelMacSch-PRE1996,TuSeifert-PRE2007}. An interaction between tilt order and surface was also studied in a model of membrane \cite{NELSON-POWERS-PRL-1992,NELSON-POWERS-JPIIFR-1992}. An interaction between the shape of membranes and the tilt order was studied with the renormalization group strategy. It was reported how thermal fluctuations of membranes associate with the strength of the interaction. However, a relation between the crumpling transition and the tilt order is still remained to be studied. 

\begin{figure}[htbp]
\centering
\includegraphics[width=110mm]{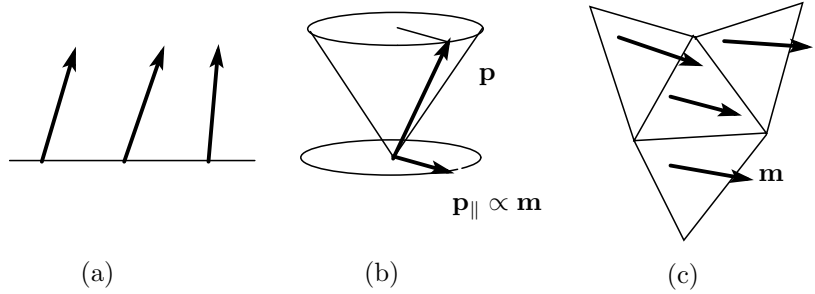}
  \caption{(a) Three-dimensional rod like structure of molecules, which is called a director, (b) the projected component ${\bf p}_\parallel$ of a director ${\bf p}$, and (c) the unit vectors ${\bf m}$ on the triangles.  The vector ${\bf m}$ is defined by ${\bf m}\!=\!{\bf p}_\parallel / |{\bf p}_\parallel | $. }
  \label{fig-1}
\end{figure}
Therefore, it is interesting to see how the crumpling transition depends on such IDOF. In order to see influences of the IDOF on the transition, we assume a three-dimensional director ${\bf p}_i$ on the triangle $i$, which is an element constructing a surface. Although IDOF of lipids cannot always simply be expressed by the tilt, we simply consider the lipid as a three dimensional vector. The directors are drawn schematically as three-dimensional vectors in Fig. \ref{fig-1}(a). The Hamiltonian corresponding to such IDOF is described by a conventional local spin-interaction between unit vectors ${\bf m}_i$, which are defined by using the projected component of ${\bf p}_{i\parallel}$ parallel to the triangle $i$ such that ${\bf m}_i\!=\!{\bf p}_{i\parallel} / |{\bf p}_{i\parallel} | $ as shown in  Fig. \ref{fig-1}(b). Figure \ref{fig-1}(c) shows the vectors ${\bf m}_i$ on triangles, which are elements of the surface.  The interaction is identical to the one of $XY$ model if the surface is flat, however, it becomes a three-dimensional one on curved surfaces in the sense that the normal perpendicular to the unit circle $S^1$ (the phase space) varies in ${\bf R}^3$. 

Our model is similar to the model in \cite{NELSON-POWERS-PRL-1992,NELSON-POWERS-JPIIFR-1992}, and therefore the interaction between the surface and the tilt order is taken into account. It is very interesting that the IDOF of molecules interacts with the external degrees of freedom that are the shape of surfaces. In many statistical systems the external degrees of freedom and the IDOF are treated independently. We know that spin models such as $XY$ model and Potts model on fluctuating surfaces were extensively studied \cite{BAILLIE-JOHNSTON-PLB-1992,CHEN-FERRENG-LANDAU-PRL-PRE-1992,JANKE-VILLANOVA-NPBSUPPL-1995,CARDY-JACOBSON-PRL-1997,JANKE-WEIGEL-APPB-2003}, however, their IDOF interact only with the intrinsic geometry and hence seems to be independent of the shape of surfaces.   

We expect that the model in this paper reveals a non-trivial influence of the tilt order on the crumpling transition. If it were not for the projection of ${\bf p}_i$ on the triangle $i$, the variable  ${\bf p}_i$ has no connection to the extrinsic geometry of surfaces. However, the projected variable ${\bf m}_i$ obviously interacts with the surface, therefore the shape of surface is influenced by correlations between ${\bf m}_i$. Conversely, we can also expect that the surface fluctuation nontrivially influences the interaction between ${\bf m}_i$. The so-called KT transition of $XY$ model on periodic flat surfaces can be changed into some other transitions or can disappear by three-dimensional effects caused by the fluctuation of surfaces or by effects of the surface topology. However, in this paper we concentrate on how the tilt order influences the crumpling transition in a broad range of interaction strength between the variables ${\bf m}_i$.

\section{Model}\label{model}

By dividing every edge of the icosahedron into $L$ pieces of uniform length, we have a triangulated surface of size $N\!=\!10L^2\!+\!2$ (= the total number of vertices). The starting configurations are thus characterized by $N_5\!=\!12$ and $N_6\!=\!N\!-\!12$, where $N_q$ is the total number of vertices with the co-ordination number $q$. 

The surface model is defined by the partition function 
\begin{eqnarray} 
\label{Part-Func}
 Z = \sum_{\bf m} \int^\prime \prod _{i=1}^{N} d X_i \exp\left[-S(X, {\cal T},{\bf m})\right],\\  
 S(X, {\cal T},{\bf m})=S_1 + b S_2 + \alpha S_3, \nonumber
\end{eqnarray} 
where $b$ is the bending rigidity, $\alpha$ is the coefficient of the $XY$ model, and $\int^\prime$ denotes that the center of the surface is fixed. $S(X,{\cal T},{\bf m})$ denotes that the Hamiltonian $S$ depends on the position variables $X$ of the vertices, the triangulation ${\cal T}$, which is fixed, and the variable {\bf m}. $\sum_{\bf m} $ denotes the summation of the IDOF corresponding to the Hamiltonian $S_3$ for the $XY$ model, which is defined by 
\begin{equation}
\label{XY} 
  S_3=\sum_{(ij)} (1-{\bf m}_i \cdot {\bf m}_j),
\end{equation} 
where $\sum_{(ij)}$ in $S_3$ is the sum over bonds $(ij)$, which are edges of the triangles $i$ and $j$. The vector ${\bf m}_i$ is defined on the triangle $i$ and a three-dimensional unit vector parallel to the triangle. As mentioned in the introduction, ${\bf m}_i$ corresponds to the projected component ${\bf p}_{i\;\parallel} $ of the director  ${\bf p}_i$; ${\bf m}_i \propto  {\bf p}_{i\; \parallel} $ and has values in the unit circle $S^1$ on the plane parallel to the triangle $i$.  The director ${\bf p}_i$ tilts a constant angle from the normal of the surface and rotates around the normal vector and interacts with the nearest neighbors. This induces the interaction described by $S_3$. Note also that $S_3$ depends on the curvature of the surface.

We emphasize that the model of Eq.(\ref{XY}) is not identical to the naive $XY$ model, whose variables have values in the unit circle $S^1$ in ${\bf R}^2$. Nevertheless, we call the model defined by $S_2$ in Eq.(\ref{XY}) as the $XY$ model as mentioned in the Introduction. The reason why we call the model as the $XY$ model is because the interaction is almost two-dimensional on a smooth two-dimensional sphere, which is locally flat.  
  
The Gaussian bond potential $S_1$, the bending energy term $S_2$ are defined by
\begin{equation}
\label{Disc-Eneg} 
  S_1=\sum_{(ij)} (X_i-X_j)^2,  \quad S_2=\sum_{(ij)} (1-{\bf n}_i \cdot {\bf n}_j),
\end{equation} 
where $\sum_{(ij)}$ in $S_1$ is the sum over bond $(ij)$ connecting the vertices $i$ and $j$, and $\sum_{(ij)}$ in $S_2$ is also the sum over bond $(ij)$, which is the common edge of the triangles $i$ and $j$. The symbol ${\bf n}_i$ in Eq. (\ref{Disc-Eneg}) denotes a unit normal vector of the triangle $i$.

\section{Monte Carlo technique}\label{MC-Techniques}
The vertices $X$ are shifted so that $X^\prime \!=\! X\!+\!\delta X$, where $\delta X$ is randomly chosen in a small sphere. The new position $X^\prime$ is accepted with the probability ${\rm Min}[1,\exp(-\Delta S)]$, where $\Delta S\!=\! S({\rm new})\!-\!S({\rm old})$. The value of Hamiltonian $S_3$ also changes due to the shift of $X_i$, since the normals of the triangles touching the vertex $i$ change.   Then, the new vector ${\bf m}_i^\prime$ is obtained by firstly projecting the old ${\bf m}_i$ to the new triangle and secondly normalizing the projected vector to the unit length. The radius of the small sphere for $\delta X$ is chosen so that the rate of acceptance for $X$ is about $50\%$. We introduce the lower bound $1\times 10^{-8}$ for the area of triangles; however, no triangle appears whose area is less than  $1\times 10^{-8}$. Therefore, we can say that no lower bound is imposed on the area of triangles. No lower bound is also imposed on the bond length. We call a sequential $N$ updates of $X$ as one Monte Carlo sweep (MCS).  

The vector ${\bf m}_i$ has its value in a unit circle, whose normal is parallel to ${\bf n}_i$ a normal vector of the triangle $i$. The new vector ${\bf m}_i^\prime$ is randomly chosen in the circle;  ${\bf m}_i^\prime$ is independent of the old  ${\bf m}_i$. As a consequence, we have about $60\% \sim 70\%$ acceptance rate for the random shift of ${\bf m}_i$. The variable ${\bf m}_i$ can be updated even when $X$ is updated, because the updates of $X$ change the normal vectors of triangles and hence ${\bf m}_i$.

 Convergence speed of MC simulations for the variables ${\bf m}_i$ is very fast compared to that for $X$, because the phase space of ${\bf m}_i$ is compact (a circle $S^1(\subset {\bf R}^3)$ ) whereas that of $X$ is noncompact (${\bf R}^3$). Therefore, the update of ${\bf m}_i$ is performed at every $10^3$ MCS in this paper. We consider that ${\bf m}_i$ should be updated after the surface shape is deformed to some extent.

We use surfaces of size $N\!=\!2562$, $N\!=\!4812$, $N\!=\!8442$  and $N\!=\!14442$. The thermalization MCS is sufficiently large: $5\times10^6 \sim 4\times 10^7$, which depends on the size of surfaces and on the values of the parameters $b$ and $\alpha$.

\section{Results}\label{Results}
\subsection{$XY$ model on rigid spheres}\label{XY-model-rigidsphere}
In this subsection we present the results of MC simulations for the $XY$ model on rigid spheres, in order to see the dependence of the correlation of spin variable ${\bf m}_i$ on the coupling constant $\alpha$. The variable $X$ of the surface is frozen in the MC simulations on the rigid sphere.

The $XY$ model is defined by the partition function
\begin{equation} 
\label{XY-model}
 Z = \sum_{\bf m}  \exp\left(-\alpha S_3\right)
\end{equation} 
 on the rigid sphere. The radius of the sphere can be chosen arbitrarily, because the partition function $Z$ in Eq.(\ref{XY-model}) depends only on the size $N$ and is independent of the radius of the sphere. 

The $XY$ model undergoes the KT transition on a flat lattice. However, we immediately understand that the long range directional order disappears on a sphere, although the directional order remains locally on the surface. As a consequence, the KT transition of the model is expected to be softened on the rigid sphere. 

\begin{figure}[htb]
\includegraphics[width=10.5cm]{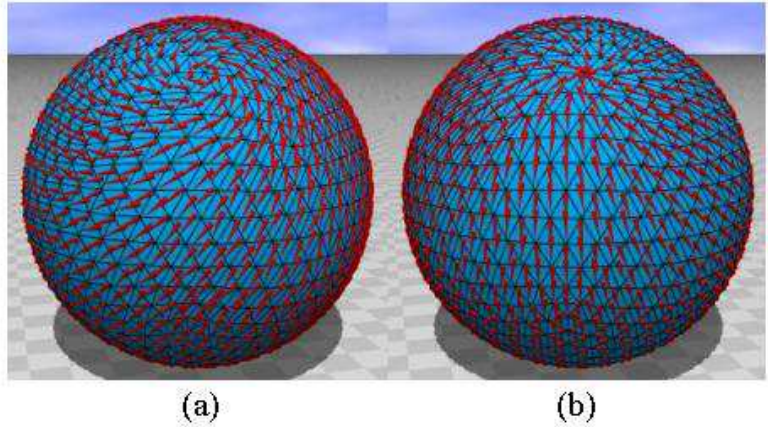}
\caption{ (Online color) (a) The KT type configuration at $\alpha\!=\!50$, and (b) the low temperature type configuration at $\alpha\!=\!200$, where small cones denote the vectors ${\bf m}$. A pair of singular points appears on both of the spheres; one of them can be seen in the figures, and the other is in the opposite side of the spheres. The surface size is $N\!=\!812$, and the magnetization $M/N_T$ of Eq.(\ref{Magnet}) is $M/N_T\!\simeq\!0.57$ in (a) and $M/N_T\!\simeq\!0.77$ in (b).} 
\label{fig-2}
\end{figure}
In order to understand the configuration at $\alpha \to \infty$, which corresponds to the zero temperature because of the unit of $\alpha$, we show in Figs. \ref{fig-2}(a) and \ref{fig-2}(b) two configurations typical to such sufficiently large $\alpha$ on the surface of size $N\!=\!812$. The configuration in Fig. \ref{fig-2}(a) obtained at $\alpha\!=\!50$ has a pair of vortices on the sphere. Flows of ${\bf m}$, denoted by cones, emerge from (go into) one vortex and go into (emerge from) the other vortex which is in the opposite side of the sphere. Two singular points do not disappear even at sufficiently large $\alpha$ because of the topological reason, in contrast to the case of periodic planar lattices. The configuration in Fig. \ref{fig-2}(b) obtained at $\alpha\!=\!200$ corresponds to the low temperature configuration. Two configurations are easily obtained by increasing $\alpha$ from small $\alpha$ such as $\alpha\!=\!5$ step by step in the simulations. The configuration of Fig. \ref{fig-2}(b) is almost stable, while that of Fig. \ref{fig-2}(a) is unstable and changes to/from that of Fig. \ref{fig-2}(b).

We consider that the singular points of ${\bf m}$ on the sphere have no influence on the KT transition. In fact, the low temperature configuration has no vortex as we see in Fig. \ref{fig-2}(b), and $XY$ model on the sphere is reported to have KT transition although the variables ${\bf m}$ have values in the circle $S^1(\subset {\bf R}^2)$ in contrast to the model in this paper \cite{BAILLIE-JOHNSTON-PLB-1992}. We should also comment on the problem of frustration, which in general appears to influence the phase structure of models defined on triangular lattices. However, the model in this paper is defined on the lattice that allows no frustrated configuration. In fact, the variables ${\bf m}$ are defined on the plaquettes (= the faces of triangles), and therefore the interaction between ${\bf m}$ forms the hexagonal (or pentagonal) lattices, which are the so-called the dual lattice. Therefore, we expect that the results are not influenced by the frustration.

\begin{figure}[htb]
\includegraphics[width=12.5cm]{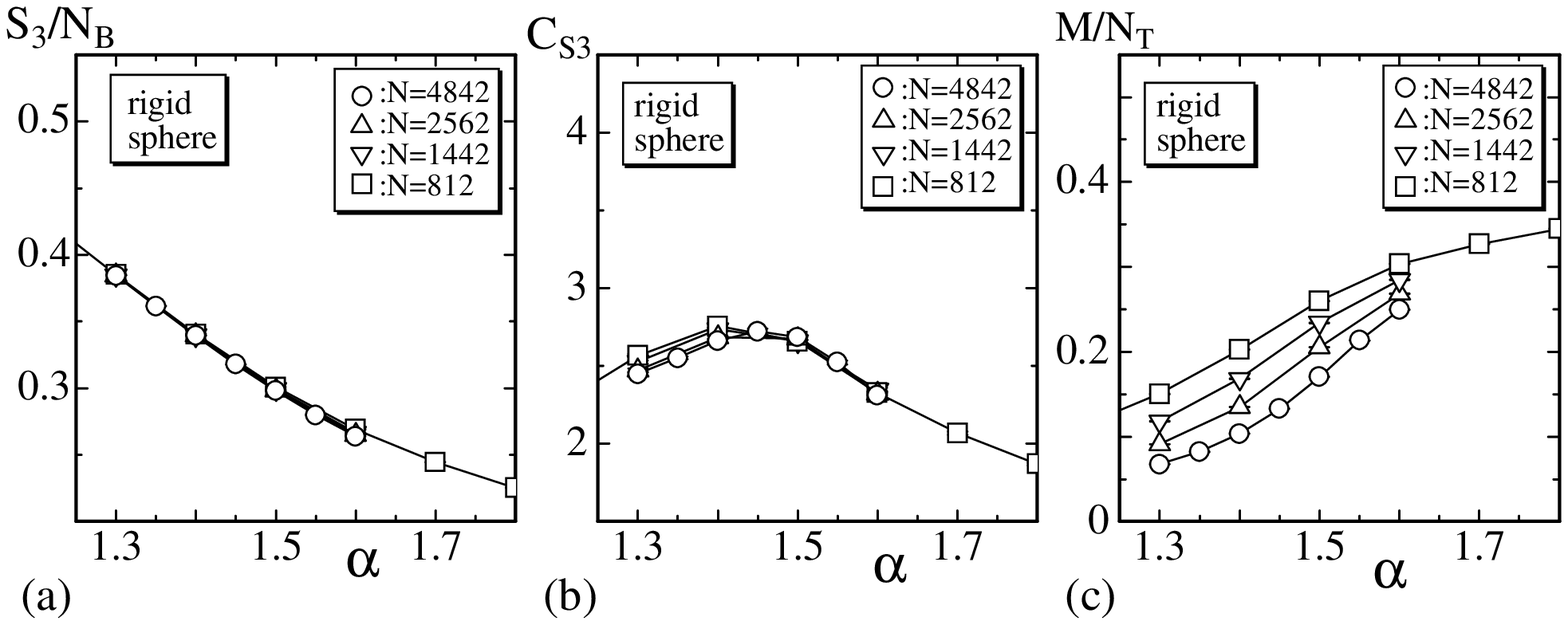}
\caption{ (a) The internal energy $S_3/N_B$ vs. $\alpha$, (b) the specific heat $C_{S_3}$ vs. $\alpha$, and the magnetization $M/N_T$ vs. $\alpha$. The model is the pure $XY$ model defined by Eq.(\ref{XY-model}) on rigid spheres. $C_{S_3}$ has the peak at $\alpha\!\simeq\!1.45$. $N_B$ and $N_T$ denote the total number of bonds and the total number of triangles, respectively. } 
\label{fig-3}
\end{figure}
Figure \ref{fig-3}(a) shows $S_3/N_B$ against $\alpha$ over the range of couplings spanning the critical region of the KT type transition. The symbols $N_B(\!=\!3N\!-\!6)$ and $N_T(\!=\!2N\!-\!4)$ denote the total number of bonds and the total number of triangles, respectively. Figure \ref{fig-3}(b) is the specific heat $C_{S_3}$ for $S_3$, which is defined by
\begin{equation} 
\label{CS3}
C_{S_3} = {\alpha^2\over N} \langle \; \left( S_3 \!-\! \langle S_3 \rangle\right)^2\rangle.
\end{equation} 
The size of surfaces is assumed as $N\!=\!812 \sim N\!=\!4842$ in the simulations on the rigid spheres. 

We find in Fig. \ref{fig-3}(b) that $C_{S_3}$ has a peak at $\alpha\!\simeq\!1.45$, where the peak value remains almost constant as $N$ increases. This implies that the model undergoes a higher order transition or the KT transition just like the $XY$ model on the flat regular lattice. The point to note is that ${\bf m}_i$ variables become relatively ordered at $\alpha\!>\!1.45$ and relatively disordered at $\alpha\!<\!1.45$ on the rigid spheres.  

The magnetization $M$ is defined by
\begin{equation} 
\label{Magnet}
 M =  \|{\bf M}\|,\quad {\bf M}=\sum_i{\bf m_i},
\end{equation} 
where $\sum_i$ is the sum over the triangle $i$. Figure \ref{fig-3}(c) shows $M/N_T$ versus $\alpha$. 

As we have seen in the low temperature configuration in Fig.\ref{fig-2}(b), $M$ is expected to be large at sufficiently large $\alpha$ even on the rigid sphere. For this reason, we expect that $M$ reflects ordering/disordering of ${\bf m}$ although ${\bf m}$ is defined parallel to the spherical surface. On fluctuating surfaces including those in the collapsed phase, $M$ is expected to have a role of order parameter of the transition. 
 
The variance $\chi_M$ of $M$ can also be defined by
$\chi_M = \langle \; \left( M \!-\! \langle M \rangle\right)^2\rangle / N_T$. 
 The KT transition in the $XY$ model on the flat lattice is the one that is characterized by the divergence of $\chi_M$ at the transition point. It is interesting to see whether the KT transition persists in the model of this paper just as in the XY model in \cite{BAILLIE-JOHNSTON-PLB-1992}. However, we do not go into this point further as mentioned in the last of the introduction. 

\subsection{Collapsing transition}\label{collapsing-transition}
In the following we concentrate on the influence of IDOF on the phase structure of the surface model. In order to do this, the coupling constant $\alpha$ is fixed to $\alpha\!=\!2$,  $\alpha\!=\!1$, and  $\alpha\!=\!0.5$. Since the transition point is $\alpha\!\sim\!1.45$ on rigid spheres, we expect that the value of $\alpha\!=\!2$ is sufficiently large for the ${\bf m}$ variables to tend to align themselves even on fluctuating spheres. On the contrary, the values $\alpha\!=\!1$ and $\alpha\!=\!0.5$ are considered to be sufficiently small to disorder the ${\bf m}$ variables.

\begin{figure}[htb]
\includegraphics[width=10.5cm]{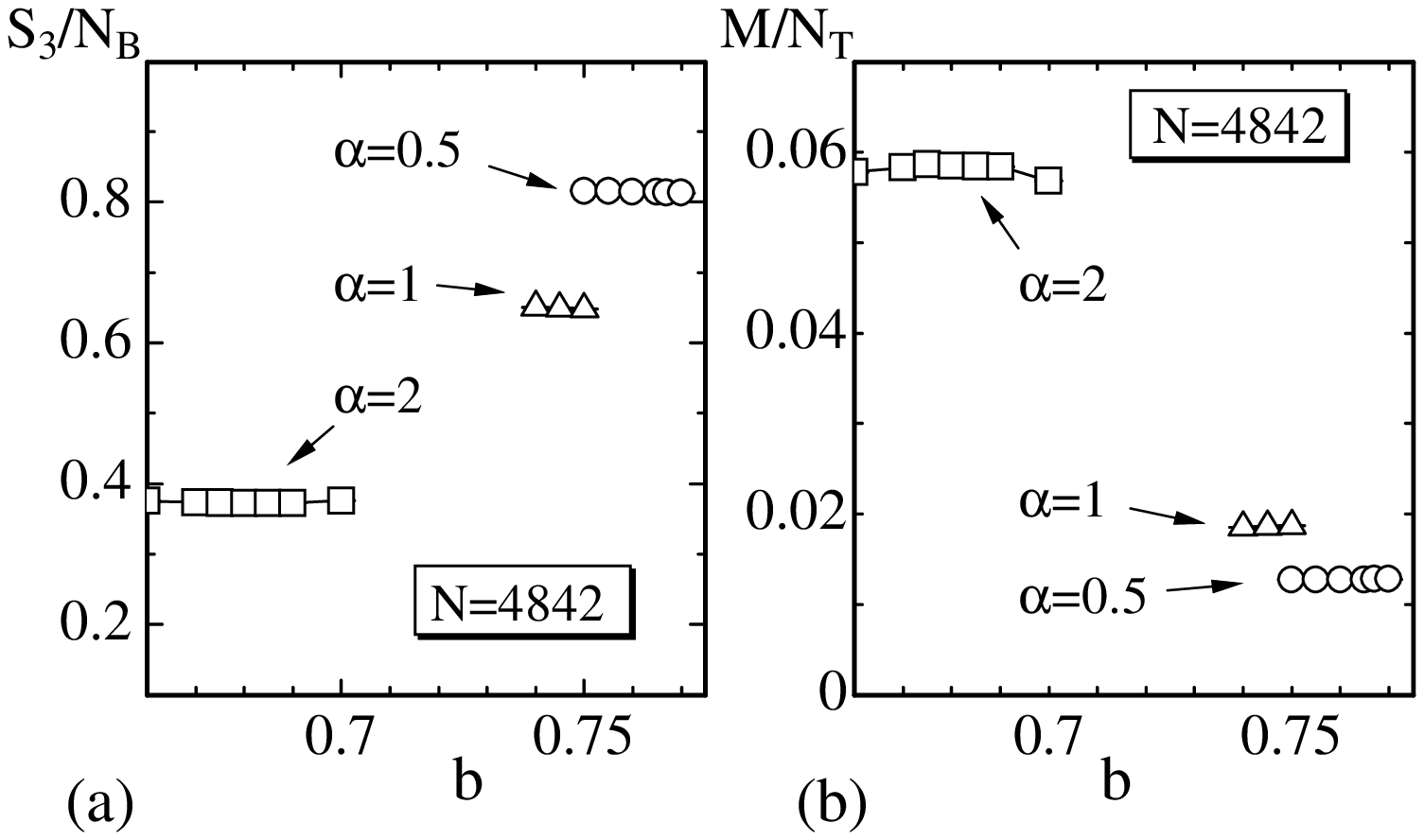}
\caption{(a) The internal energy $S_3/N_B$ versus $b$, and (b) the magnetization $M/N_T$ versus $b$, which were obtained at $\alpha\!=\!2$,  $\alpha\!=\!1$, and $\alpha\!=\!0.5$.  The model is the one defined by Eq.(\ref{Part-Func}). The range of assumed $b$ is spanning the collapsing transition point in each $\alpha$.  } 
\label{fig-4}
\end{figure}

Before analyzing the collapsing transition, we show some of the quantities such as the magnetization $M/N_T$ and the internal energy $S_3/N_B$ to get information on the configurations corresponding to the ${\bf m}$ variables. Figures \ref{fig-4}(a) and \ref{fig-4}(b) show $S_3/N_B$ and $M/N_T$ versus the bending rigidity $b$ obtained under the conditions $\alpha\!=\!2$,  $\alpha\!=\!1$, and  $\alpha\!=\!0.5$. The data are obtained on fluctuating spheres. The range of $b$ in each $\alpha$ is the region of collapsing transition point $b_c$, where the surface collapses at $b\!<\!b_c$ and swells at $b\!>\!b_c$. 

The value $S_3/N_B\simeq 0.37$ at $\alpha\!=\!2$ in Fig.\ref{fig-4}(a) indicates that the interaction strength $\alpha\!=\!2$ corresponds to the ordered phase of the $XY$ model on rigid sphere, which was shown in Figs.\ref{fig-3}(a) and  \ref{fig-3}(b). We see also in Fig.\ref{fig-4}(a) that the vectors ${\bf m}_i$ are almost decorrelated to each other at $\alpha\!=\!1$ and $\alpha\!=\!0.5$. We get the same information on the configurations of ${\bf m}_i$ from $M/N_T$ in Fig.\ref{fig-4}(c) by comparing the values of $M/N_T$ to the corresponding ones in Fig.\ref{fig-3}(c). 

We find also from Figs.\ref{fig-4}(a) and \ref{fig-4}(b) that both $M/N_T$ and $S_3/N_B$ are almost independent of $b$. This implies that the spin vectors ${\bf m}$ are hardly influenced by whether the surface is smooth or collapsed even though ${\bf m}_i$ is constrained to be parallel to the triangle $i$.

\begin{figure}[htb]
\includegraphics[width=12.5cm]{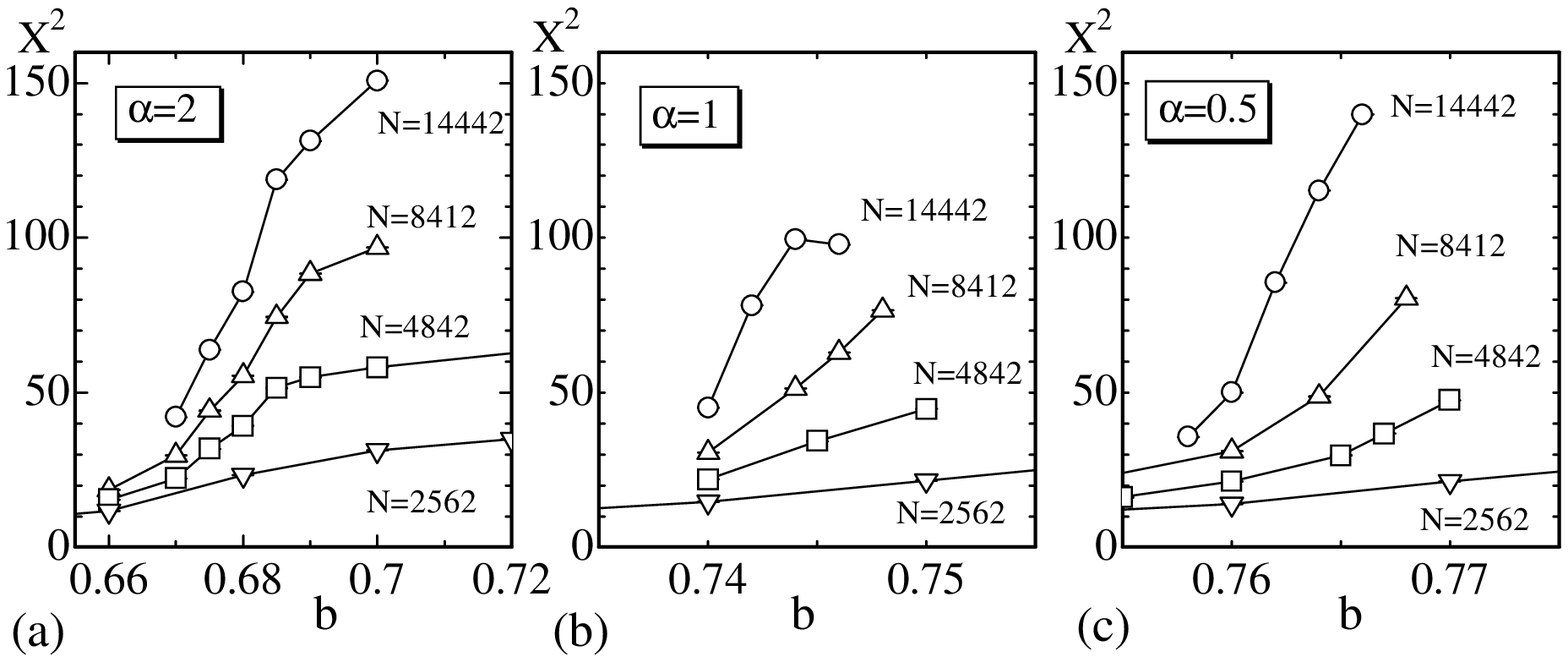}
\caption{The internal energy $S_3/N_B$ versus $b$ at (a) $\alpha\!=\!2$,  (b) $\alpha\!=\!1$, and  (c) $\alpha\!=\!0.5$. } 
\label{fig-5}
\end{figure}
Now, let us turn to the collapsing transition. We show how the IDOF influences the collapsing transition. 
The mean square size $X^2$ is defined by
\begin{equation}
\label{X2}
X^2={1\over N} \sum_i \left(X_i-\bar X\right)^2, \quad \bar X={1\over N} \sum_i X_i,
\end{equation}
where $\bar X$ is the center of the surface, and is plotted in Figs. \ref{fig-5}(a)--(c) against $b$. $X^2$ in the figure was obtained at $\alpha\!=\!2$, $\alpha\!=\!1$, and $\alpha\!=\!0.5$. We find that the transition point $b_c$, where $X^2$ rapidly varies, moves right on the $b$ axis with decreasing $\alpha$. In the limit of $\alpha\!=\!0$, the transition point should be $b_c\to b_c^0$, where $b_c^0\!\simeq \!0.77$ \cite{KOIB-PRE-2005} is the transition point of the model without the IDOF. This implies that the surface is softened by the interaction between the surface and the IDOF, because the stiffness of the surface is considered to be reduced if the transition point $b_c$ decreases even though the parameter $b$ is itself not always identical to the macroscopic bending rigidity of the surface.

\begin{figure}[htb]
\includegraphics[width=12.5cm]{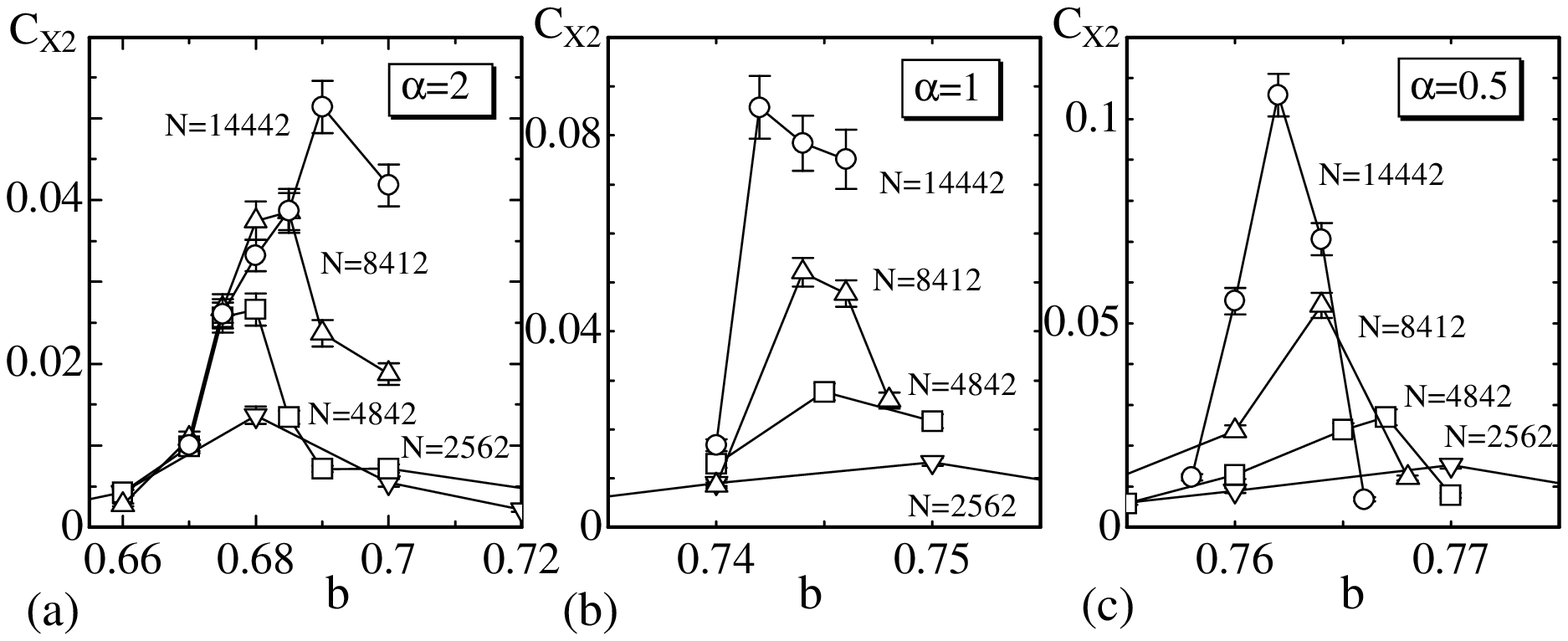}
\caption{The specific heat $C_{S_3}$ versus $b$ at (a) $\alpha\!=\!2$,  (b) $\alpha\!=\!1$, and  (c) $\alpha\!=\!0.5$. Anomalous peaks indicate a collapsing transition between the smooth spherical phase and a collapsed phase.} 
\label{fig-6}
\end{figure}
The fluctuation of $X^2$ is defined by 
\begin{equation} 
\label{CX2}
C_{X^2} = {1\over N} \langle \; \left( X^2 \!-\! \langle X^2 \rangle\right)^2\rangle,
\end{equation} 
which is expected to reflect how large the surface size fluctuates. Figures \ref{fig-6}(a)--(c) show $C_{X^2}$ against $b$ at $\alpha\!=\!2$, $\alpha\!=\!1$, and $\alpha\!=\!0.5$. As expected from Figs.\ref{fig-5}(a)--(c), the peak position of $C_{X^2}$ moves to the right on the $b$ axis as $\alpha$ decreases from $\alpha\!=\!2$ to $\alpha\!=\!0.5$. We find also that the peak value $C_{X^2}^{\rm max}$ itself increases with decreasing $\alpha$. This second observation implies that the shape fluctuation is slightly suppressed by the interaction between the surface and the IDOF, because $C_{X^2}^{\rm max}$ decreases with increasing $\alpha$. 

\begin{figure}[htb]
\includegraphics[width=12.5cm]{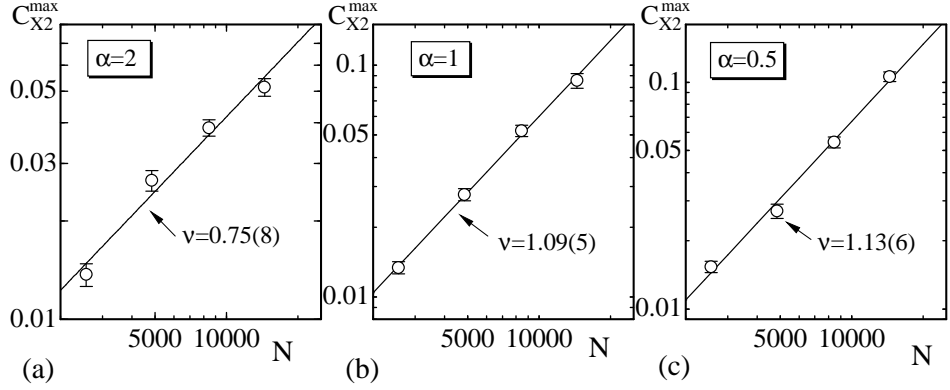}
\caption{Log-log plots of the peak value $C_{S_3}^{\rm max}$ against $N$ at (a) $\alpha\!=\!2$,  (b) $\alpha\!=\!1$, and  (c) $\alpha\!=\!0.5$. The straight lines were drawn by fitting the data to Eq.(\ref{mu-scale}). } 
\label{fig-7}
\end{figure}
In order to see the order of the collapsing transition, we show log-log plots of $C_{X^2}^{\rm max}$ against $N$ in Figs. \ref{fig-7}(a)--(c). The straight lines are drawn by fitting the data to
\begin{equation} 
\label{mu-scale}
C_{X^2}^{\rm max} \sim N^{\nu},
\end{equation} 
where $\nu$ is a critical exponent of the collapsing transition. Thus we have
\begin{eqnarray} 
\label{mu-value}
\nu_{\alpha=2}=0.75\pm 0.08\quad (\alpha=2), \nonumber \\
\quad\nu_{\alpha=1}=1.09\pm 0.05\quad (\alpha=1), \\
\quad\nu_{\alpha=0.5}=1.13\pm 0.06\quad (\alpha=0.5). \nonumber
\end{eqnarray} 
From the finite-size scaling (FSS) theory, the second and the third results in Eq.(\ref{mu-value}) indicate that the transition is of first-order, because $\nu$ are compatible to $\nu\!=\!1$. On the contrary, the first result $\nu_{\alpha=2}\!=\!0.75(8)$ slightly deviates from  $\nu\!=\!1$, as a consequence, the FSS analysis of $C_{X^2}^{\rm max}$ fails to predict that the transition is discontinuous at $\alpha\!=\!2$.

\begin{figure}[htb]
\includegraphics[width=12.5cm]{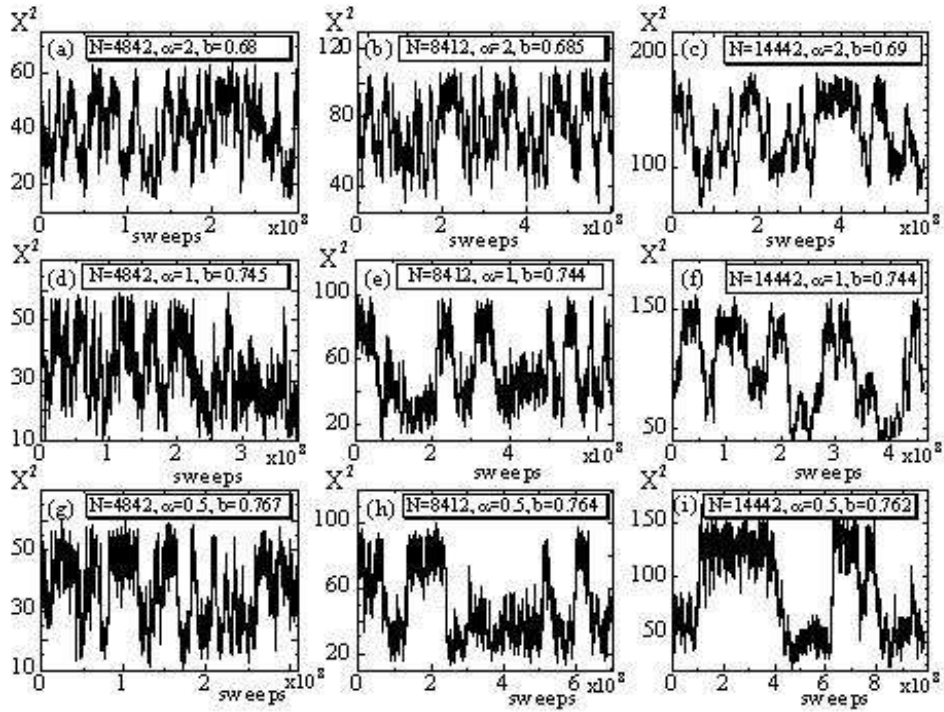}
\caption{The variation of $X^2$ against MCS at (a)--(c) $\alpha\!=\!2$,  (d)--(f) $\alpha\!=\!1$, and (g)--(i) $\alpha\!=\!0.5$. The data were obtained at the transition point $b_c$, which depends on both $N$ and $\alpha$. } 
\label{fig-8}
\end{figure}
The surface shape is expected to be discontinuously changed at the transition point. Then, it is natural to ask what is the Hausdorff dimension $H$ of the surface at the smooth phase and at the collapsed phase. We show in Figs.\ref{fig-8}(a)--(i) the variation of $X$ against MCS obtained at the transition point $b_c$ with $\alpha\!=\!2$, $\alpha\!=\!1$, and $\alpha\!=\!0.5$. It is almost clear that $X^2$ discontinuously changes between the smooth phase and the collapsed phase in almost all cases shown in Figs.\ref{fig-8}(a)--(i).

\begin{figure}[htb]
\includegraphics[width=12.5cm]{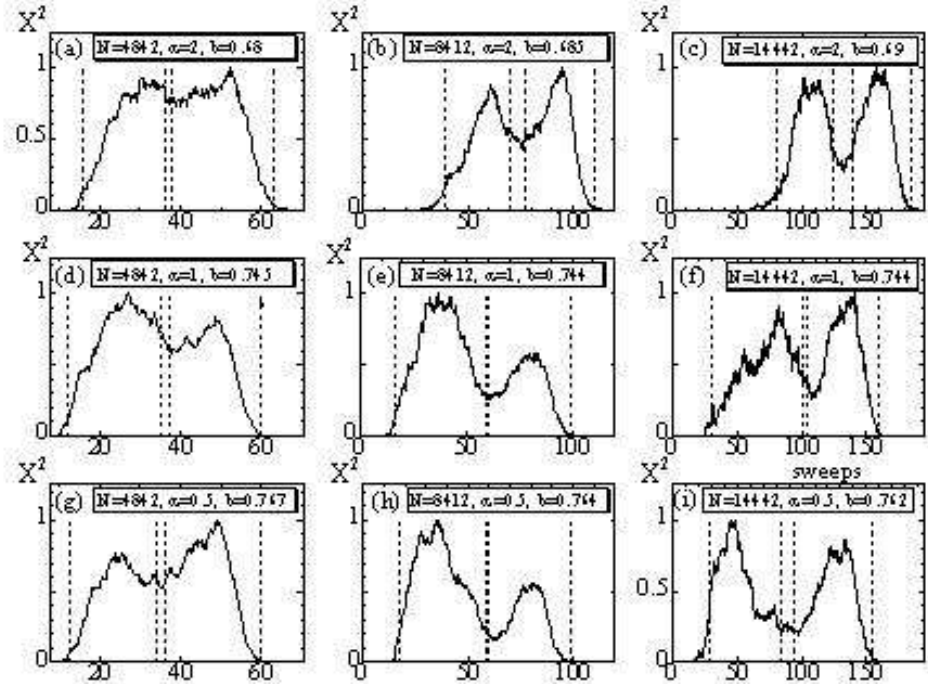}
\caption{Normalized histograms $h(X^2)$ for the distribution of $X^2$ obtained  at (a)--(c) $\alpha\!=\!2$,  (d)--(f) $\alpha\!=\!1$, and (g)--(i) $\alpha\!=\!0.5$. Those $h(X^2)$ correspond to the variations of $X^2$ shown in Figs.\ref{fig-8}(a)--(i). The dashed vertical lines denote $X^{2 \;{\rm col}}_{\rm min}$, $X^{2 \;{\rm col}}_{\rm max}$ and $X^{2 \;{\rm smo}}_{\rm min}$, $X^{2 \;{\rm smo}}_{\rm max}$, which will be shown in Table \ref{table-1}.  } 
\label{fig-9}
\end{figure}
Figures \ref{fig-9}(a)--(i) show the normalized histogram $h(X^2)$ of variation $X^2$, which was shown in Figs.\ref{fig-8}(a)--(i). A double peak structure is clearly seen in $h(X^2)$ at all values of $\alpha$ except on the surface of size $N\!=\!4842$ in Fig.\ref{fig-9}(a). The reason of this is only the size effect; $h(X^2)$ has the double peak on sufficiently large sized surfaces. Thus, the collapsing transition is confirmed to be first-order even at $\alpha\!=\!2$, where the first-order transition was not confirmed by the FSS analysis for $C_{X^2}^{\rm max}$ in Eq.(\ref{mu-scale}).

\begin{table}[hbt]
\caption{ The lower bound $X^{2 \;{\rm col}}_{\rm min}$ and the upper bound $X^{2 \;{\rm col}}_{\rm max}$ in the collapsed state, and the lower bound $X^{2 \;{\rm smo}}_{\rm min}$ and the upper bound $X^{2 \;{\rm smo}}_{\rm max}$ for obtaining the mean value $X^2({\rm smo})$ in the smooth state. }
\label{table-1}
\begin{center}
 \begin{tabular}{cccccc}
$\alpha$  & $N$ & $X^{2 \;{\rm col}}_{\rm min}$ & $X^{2 \;{\rm col}}_{\rm max}$ & $X^{2 \;{\rm smo}}_{\rm min}$ & $X^{2 \;{\rm smo}}_{\rm max}$   \\
 \hline
  2  & 2562 & 12 & 22  & 23 & 36  \\
  2  & 4842 & 16 & 36  & 38 & 63  \\
  2  & 8412 & 39 & 70  & 77 & 110  \\
  2  & 14442 & 80 & 125 & 140 & 185  \\
 \hline
  1  & 2562 & 13 & 21  & 22 & 34  \\
  1  & 4842 & 12 & 35  & 37 & 60  \\
  1  & 8412 & 17 & 59  & 60 & 99  \\
  1  & 14442 & 30 & 100 & 105 & 160  \\
 \hline
  0.5  & 2562 & 8  & 20  & 21 & 33  \\
  0.5  & 4842 & 13 & 35  & 36 & 60  \\
  0.5  & 8412 & 17 & 59  & 60 & 99  \\
  0.5  & 14442 & 50 & 90  & 100 & 160  \\
 \hline
 \end{tabular} 
\end{center}
\end{table}
The double peak in $h(X^2)$ allows us to calculate the mean value $X^{2\; {\rm smo}}$ of $X^2$ in the smooth phase and the mean value $X^{2\; {\rm col}}$ in the collapsed phase. These mean values can be calculated by assuming the lower (upper) bound $X^2_{\rm min}$ ($X^2_{\rm max}$ ) and by averaging $X^2$ in the range $X^{2 \;{\rm col(smo)}}_{\rm min}< X^2 < X^{2 \;{\rm col(smo)}}_{\rm max}$, which includes one of the two peaks. The dashed lines drawn vertically in Figs.\ref{fig-9}(a)--(i) denote the lower and upper bounds; four dashed lines in each figure respectively correspond to  $X^{2 \;{\rm col}}_{\rm min}$, $X^{2 \;{\rm col}}_{\rm max}$, $X^{2 \;{\rm smo}}_{\rm min}$, and $X^{2 \;{\rm smo}}_{\rm max}$. Table \ref{table-1} show the lower and the upper bounds $X^2_{\rm min}$, $X^2_{\rm max}$ including those shown as dashed lines in Figs.\ref{fig-9}(a)--(i).

\begin{figure}[htb]
\includegraphics[width=12.5cm]{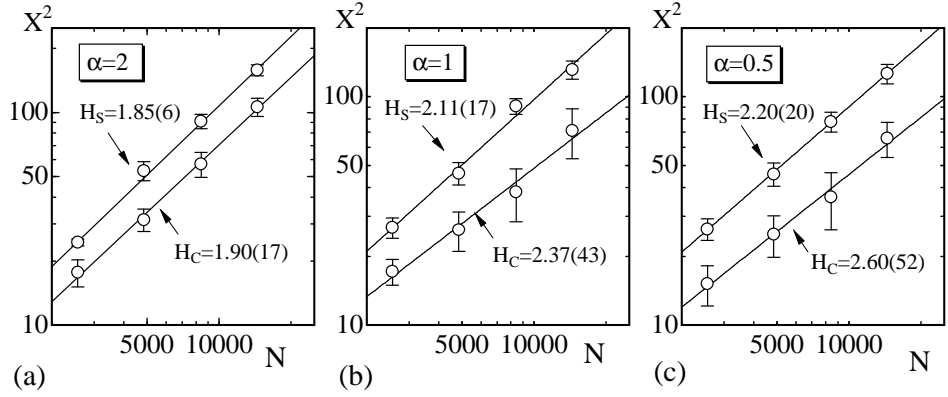}
\caption{Log-log plots of the mean values $X^2$ against $N$ obtained in the smooth phase and the collapsed phase at (a) $\alpha\!=\!2$,  (b) $\alpha\!=\!1$, and  (c) $\alpha\!=\!0.5$. The mean values $X^2$ were obtained by averaging $X^2$ between the lower bound $X^2_{\rm min}$ and the upper bound $X^2_{\rm max}$, which are shown in Table \ref{table-1} and also indicated by vertical dashed lines in Figs.\ref{fig-9}(a)--(i).   } 
\label{fig-10}
\end{figure}
Figures \ref{fig-10}(a)--(c) show $X^2$ versus $N$ in a log-log scale, where $X^2$ were obtained by averaging the data between the lower and the upper bounds listed in Table \ref{table-1}. The error bars in Figs.\ref{fig-10}(a)--(c) are the standard deviations. The straight lines were drawn by the power fit of the form
\begin{equation} 
\label{Hausdorff-fitting}
X^2 \sim N^{2/H},
\end{equation} 
where $H$ is the Hausdorff dimension.  Then we have

\begin{eqnarray} 
\label{H-results}
H_{\rm smo} = 1.85 \pm 0.06,  \quad H_{\rm col} = 1.90 \pm 0.17 \; (\alpha\!=\!2 ),\nonumber \\
H_{\rm smo} = 2.11 \pm 0.17,  \quad H_{\rm col} = 2.37 \pm 0.43 \; (\alpha\!=\!1 ), \\
H_{\rm smo} = 2.20 \pm 0.20,  \quad H_{\rm col} = 2.60 \pm 0.52 \; (\alpha\!=\!0.5 ). \nonumber
\end{eqnarray} 
The results $H_{\rm smo}$ in the smooth phase are almost identical to the expected value $H\!=\!2$, which is the topological dimension of surfaces. We see that the results $H_{\rm col}$ in the collapsed phase obtained at $\alpha\!=\!2$ and $\alpha\!=\!1$ are also almost identical to $H\!=\!2$, and that $H_{\rm col}$ at $\alpha\!=\!0.5$ remains in the physical bound $H\!=\!3$. In the limit of $\alpha\to 0$, $H_{\rm col}$ is expected to be $H\!=\!2.59(57)$, which was obtained in the case of $\alpha\!=\!0$ in \cite{KOIB-PRE-2005}. The result $H_{\rm col}$ at $\alpha\!=\!0.5$ in Eq.(\ref{H-results}) is almost identical to this value. Not only the smooth phase but also the collapsed phase is therefore considered to be unaffected in the presence of the IDOF in the range of small to medium $\alpha$.   

\subsection{Surface fluctuations}\label{surface-fluctuation}
\begin{figure}[hb]
\includegraphics[width=12.5cm]{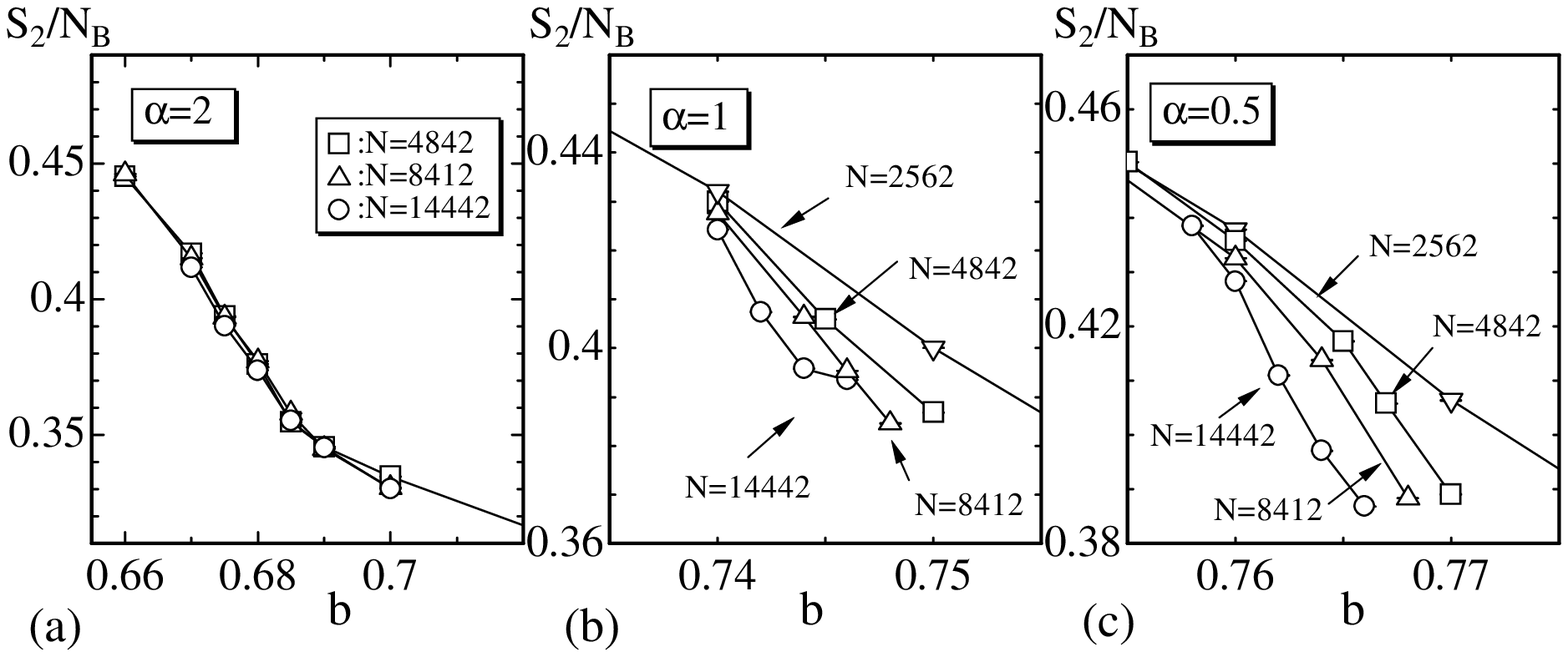}
\caption{The bending energy $S_2/N_B$ versus $b$ at (a) $\alpha\!=\!2$,  (b) $\alpha\!=\!1$, and  (c) $\alpha\!=\!0.5$, where $N_B$ is the total number of bonds. } 
\label{fig-11}
\end{figure}
The bending energy $S_2$ can reflect how smooth the surface is, and therefore we plot $S_2/N_B$ versus $b$ in Figs.\ref{fig-11}(a)--(c), where $N_B$ is the total number of bonds. The results in Fig.\ref{fig-11}(a) at $\alpha\!=\!2$ are independent of $N$ and hence indicate that the surface fluctuation is suppressed and the phase transition disappears. On the contrary, $S_2/N_B$ shown in Figs.\ref{fig-11}(a) and \ref{fig-11}(b) varies rapidly with increasing $N$, and this is considered to be a signal of phase transition although a discontinuity can not be seen in those $S_2/N_B$.

\begin{figure}[htb]
\includegraphics[width=12.5cm]{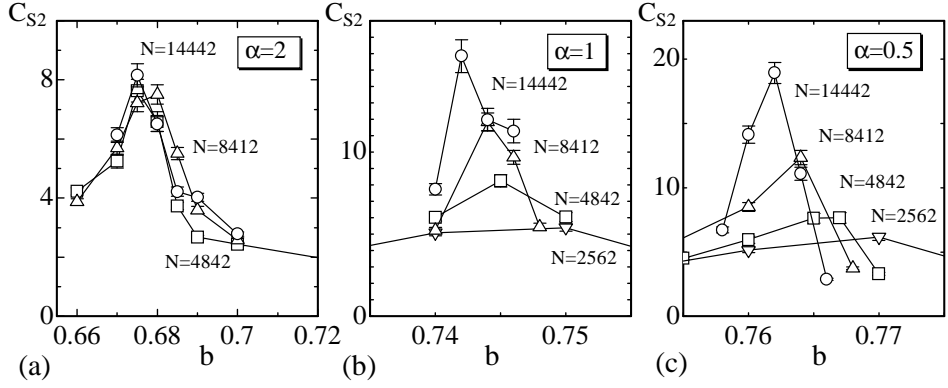}
\caption{The specific heat $C_{S_2}$ versus $b$ at (a) $\alpha\!=\!2$,  (b) $\alpha\!=\!1$, and  (c) $\alpha\!=\!0.5$.  Anomalous peaks $C_{S_2}^{\rm max}$ indicate the existence of phase transition for surface fluctuations.  } 
\label{fig-12}
\end{figure}
In order to see the order of the transition more clearly, we plot in Figs.\ref{fig-12}(a)--(c) the specific heat $C_{S_2}$ for the bending energy $S_2$, which is defined by
\begin{equation} 
\label{CS2}
C_{S_2} = {b^2\over N} \langle \; \left( S_2 \!-\! \langle S_2 \rangle\right)^2\rangle.
\end{equation} 
The anomalous behavior seen in $C_{S_2}$ indicates a phase transition between the smooth phase and the collapsed phase. However, as we see in Fig.\ref{fig-12}(a), the peak value $C_{S_2}^{\rm max}$ remains constant even when $N$ increases. This is consistent to the behavior of $S_2/N_B$ in Fig.\ref{fig-11}(a).

\begin{figure}[htb]
\includegraphics[width=12.5cm]{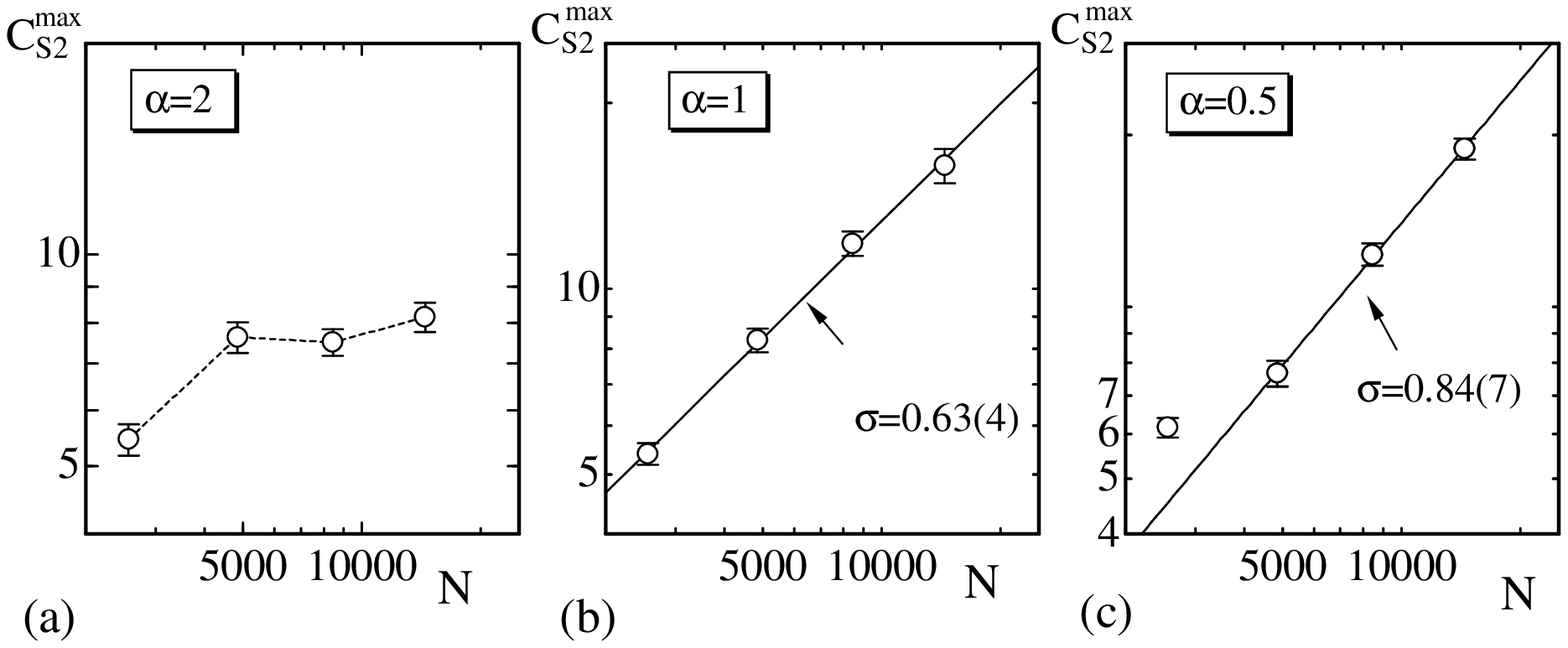}
\caption{Log-log plots of $C_{S_2}^{\rm max}$ against $N$ at (a) $\alpha\!=\!2$,  (b) $\alpha\!=\!1$, and  (c) $\alpha\!=\!0.5$. The straight lines in (b) and (c) were drawn by fitting the data (the largest three data in (c)) to the form of Eq.(\ref{sigma-scale}).} 
\label{fig-13}
\end{figure}
 The peak value $C_{S_2}^{\rm max}$ of $C_{S_2}$ is plotted against $N$ in a log-log scale in Figs.\ref{fig-13}(a)--(c).  It is apparent that $C_{S_2}^{\rm max}$ in Fig.\ref{fig-13}(a) stops growing with increasing $N$, which is in sharp contrast to the cases in  Figs.\ref{fig-13}(b),(c). The straight lines in Figs.\ref{fig-13}(b) and \ref{fig-13}(c) are drawn by fitting the data to the form
\begin{equation} 
\label{sigma-scale}
C_{S_2}^{\rm max} \sim N^\sigma,
\end{equation} 
where $\sigma$ is a critical exponent of the transition. The largest three data are used in the fitting in the case of $\alpha\!=\!0.5$ in Figs.\ref{fig-13}(c). Thus, we have
\begin{eqnarray} 
\label{sigma-value}
\sigma = 0.63 \pm 0.04\quad (\alpha=1), \nonumber \\
 \sigma = 0.84 \pm 0.07\quad (\alpha=0.5). 
\end{eqnarray} 
The first result in Eq.(\ref{sigma-value}) implies that the transition is of second order (or continuous) at $\alpha\!=\!1$, because $\sigma$ is considered to be $\sigma \!<\! 1$. From the finite-size scaling (FSS) theory, we know that $\sigma \!<\! 1$ corresponds a continuous transition. On the contrary, the transition appear to be discontinuous at $\alpha\!=\!0.5$, because the second result in Eq.(\ref{sigma-value}) is considered to be almost equal to $\sigma \!=\! 1$. We expect that the order of transition turns to be discontinuous when $\alpha$ is reduced, because a discontinuous transition can be seen in the case of $\alpha\!=\!0$ \cite{KOIB-PRE-2005}. 

\begin{figure}[htb]
\includegraphics[width=12.5cm]{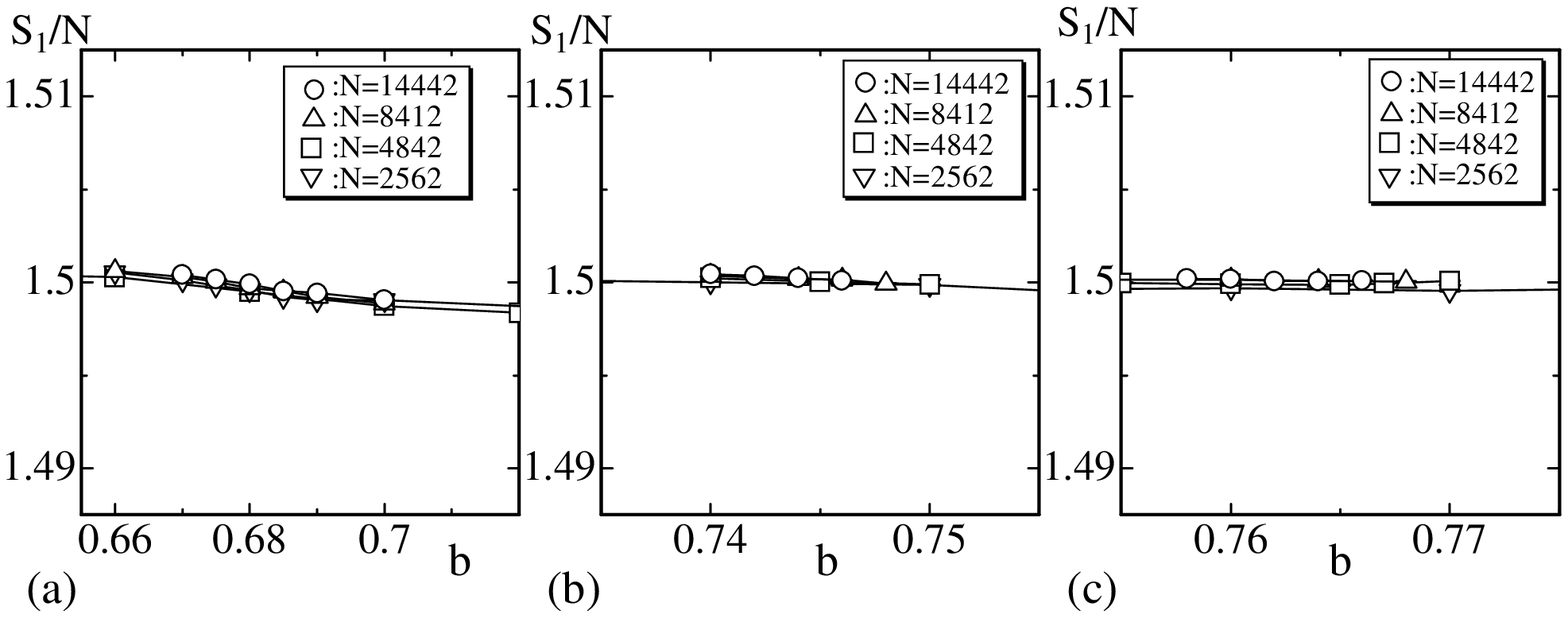}
\caption{The Gaussian bond potential $S_1/N$ versus $b$ at (a) $\alpha\!=\!2$,  (b) $\alpha\!=\!1$, and  (c) $\alpha\!=\!0.5$. The expected relation $S_1/N\!\simeq\!3/2$ is satisfied. } 
\label{fig-14}
\end{figure}
Finally, we show $S_1/N$ versus $b$ in Figs.\ref{fig-14}(a)--(c). $S_1/N$ is expected to be $S_1/N\!\simeq\!3/2$ because of the scale invariant property of the partition function. Consequently, we can use this relation to see whether the simulations are performed successfully. It is easy to see that the relation is satisfied in almost all cases except in the smooth phase at $\alpha\!=\!2$; however, the deviation is still very small compared to the value $S_1/N$. Therefore, we consider that the simulations are performed correctly.

\section{Summary and Conclusion}\label{Conclusion}
We have investigated an interaction between the tilt order and the shape of surfaces of the conventional surface model of Helfrich, Polyakov and Kleinert by Monte Carlo simulations on triangulated spherical lattices. The purpose of this study is to see how the tilt order influences the collapsing transition and the surface fluctuation transition, both of which were reported to be of first-order in the conventional surface model \cite{KOIB-PRE-2005}. The Hamiltonian of the model in this paper is defined by a linear combination of the HPK Hamiltonian and that of the $XY$ model. The unit vector ${\bf m}_i$ of the $XY$ model is defined on the triangle $i$ by projecting a three-dimensional vector on the triangle, where the three-dimensional vector is assumed to represent a lipid molecule usually called a director.  Since the vector ${\bf m}_i$ is parallel to the triangle $i$, the $XY$ model is not identical to the naive $XY$ model defined on the planar lattices. The parameter $\alpha$, which is the coefficient of the $XY$ Hamiltonian, is assumed to $\alpha\!=\!2$,  $\alpha\!=\!1$, and $\alpha\!=\!0.5$ in the simulations. In the case of $\alpha\!=\!2$ the vectors ${\bf m}$ are considered to be in a relatively ordered state, while in the cases $\alpha\!=\!1$ and $\alpha\!=\!0.5$ they are in disordered states. 

We find that the variables ${\bf m}_i$ change depending only on $\alpha$ and are almost independent of whether the surface is smooth or not. In fact, the internal energy $S_3/N_B$ and the magnetization $M/N_T$ remain almost constant in the range of $b$ including $b_c$ the collapsing transition point.    

It is also observed that the collapsing transition is not so strongly influenced by the tilt order. The transition is slightly softened in the presence of tilt order, however, it remains in first-order and occurs almost independent of $\alpha$ at least up to $\alpha\!=\!2$. Furthermore, the collapsing phase at the transition point is characterized by a physical Hausdorff dimension, i.e., $ H_{\rm col}\!<\!3$ in all cases $\alpha\!=\!2$,  $\alpha\!=\!1$, and $\alpha\!=\!0.5$. This result is consistent with the physical Hausdorff dimension at $\alpha\!=\!0$ reported in \cite{KOIB-PRE-2005}. 

On the other hand, the transition of surface fluctuations is influenced by the tilt order. The transition appears to remain discontinuous at $\alpha\!=\!0.5$, where the variables ${\bf m}_i$ weakly correlate with each other and are relatively at random. As the coefficient increases to $\alpha\!=\!1$, the transition is softened and turns to be a continuous one. Moreover, the transition disappears and turns to be a higher-order one as $\alpha$ increases to $\alpha\!=\!2$. This result leads us to conclude that the collapsing transition is not always accompanied by the surface fluctuation transition in the surface model with internal degrees of freedom such as the tilt order. 

Finally, we comment on whether the phase structure is influenced by the singular points of ${\bf m}$, which appear even at zero temperature due to the surface topology. The exponents $\nu$ shown in Eq.(\ref{mu-scale}) and the Hausdorff dimension $H$ in Eq.(\ref{H-results}), both of which characterize the collapsing transition, are not influenced by the singular points. In fact, the collapsing transition is almost independent of the variables ${\bf m}$ as we have shown in this paper. However, it remains unclarified whether or not the exponent $\sigma$ in Eq.(\ref{sigma-value}) is influenced by the singularity; more precisely, it remains to be studied whether or not the softening of the surface fluctuation transition is caused only by the singularity of ${\bf m}$ or the topology of surface, although KT transition itself is expected to be independent of the surface topology as discussed in Section \ref{XY-model-rigidsphere}. 

This work is supported in part by a Grant-in-Aid for Scientific Research from Japan Society for the Promotion of Science. 




\begin{thebibliography}{00}
\bibitem{NELSON-SMMS2004}
D. Nelson, in {Statistical Mechanics of Membranes and Surfaces, Second Edition}, edited by  D. Nelson, T.Piran, and S.Weinberg, (World Scientific, 2004), p.1. 

\bibitem{Gompper-Schick-PTC-1994}
G. Gompper and M. Schick, \textit{Self-assembling amphiphilic systems}, In
\textit{Phase Transitions and Critical Phenomena 16}, C. Domb and J.L. Lebowitz, Eds. (Academic Press, 1994) p.1.

\bibitem{Bowick-PREP2001}
 M. Bowick and A. Travesset, Phys. Rep. 344 (2001) 255.
 

\bibitem{HELFRICH-1973}
 W. Helfrich, Z. Naturforsch, {\bf 28}c (1973) 693.

\bibitem{POLYAKOV-NPB1986}
 A.M. Polyakov, Nucl. Phys. B {\bf 268} (1986) 406.

\bibitem{KLEINERT-PLB1986}
 H. Kleinert, Phys. Lett. B {\bf 174} (1986) 335.

\bibitem{KD-PRE2002}
J-P. Kownacki and H. T. Diep, Phys. Rev. E {\bf 66},  (2002)  066105.

\bibitem{KOIB-PRE-2004-1}
H. Koibuchi, N. Kusano, A. Nidaira, K. Suzuki, and M. Yamada, Phys. Rev. E {\bf 69}, 066139
(2004).

\bibitem{KOIB-PRE-2005}
 H. Koibuchi and T. Kuwahata, Phys. Rev. E {\bf 72}, (2005) 026124. 

\bibitem{KOIB-NPB-2006}
 I. Endo and H. Koibuchi, Nucl. Phys. B {\bf 732} [FS], (2006) 732. 

\bibitem{DAVID-SMMS2004}
F. David, in {Statistical Mechanics of Membranes and Surfaces, Second Edition}, edited by  D. Nelson, T.Piran, and S.Weinberg, (World Scientific, 2004), p.149. 

\bibitem{Peliti-Leibler-PRL1985}
 L. Peliti and S. Leibler, Phys. Rev. Lett. \textbf{54} (15), 1690 (1985).

\bibitem{DavidGuitter-EPL1988}
 F. David and E. Guitter, Europhys. Lett,  \textbf{5} (8), 709  (1988).

\bibitem{PKN-PRL1988}
M. Paczuski, M. Kardar, and D. R. Nelson, Phys. Rev. Lett. \textbf{60}, 2638 (1988).

\bibitem{KANTOR-NELSON-PRA1987}
 Y. Kantor and  D.R. Nelson, Phys. Rev. A \textbf{36}, 4020  (1987).

\bibitem{AMBJORN-NPB1993}
 J. Ambjorn, A. Irback, J. Jurkiewicz, and B. Petersson, Nucl. Phys. B \textbf{393}, 571 (1993).

\bibitem{TYY-NJP-2003}
Y. Tabe, T. Yamamoto, and H Yokoyama, New J. Phys. {bf 5} (2003) 65. 

\bibitem{LL-PRB1987}
S. Leibler and  R. Lipowsky,
Phys. Rev. B {\bf 35}, 7004 (1987).  

\bibitem{HelfrichProst-PRA1988} W. Helfrich and J. Prost, Phys. Rev. A  \textbf{38}, 3065 (1988).

\bibitem{ZJX-PRL1990} Ou-Yang Zhong-can and Liu Ji-xing, Phys. Rev. Lett. \textbf{65}, 1679 (1990).

\bibitem{SelMacSch-PRE1996} J. V. Selinger, F. C. MacKintosh and J. M. Schnur, Phys. Rev. E \textbf{53}, 3804 (1996).

\bibitem{TuSeifert-PRE2007} Z. C. Tu and U. Seifert, Phys. Rev. E \textbf{76}, 031603 (2007).

\bibitem{NELSON-POWERS-PRL-1992}
P. Nelson and T. Powers, Phys. Rev. Lett. {\bf 69} (1992) 3409.

\bibitem{NELSON-POWERS-JPIIFR-1992}
P. Nelson and T. Powers, J. Phys. II France {\bf 3} (1993) 1535. 

\bibitem{BAILLIE-JOHNSTON-PLB-1992}
C.F. Baillie and D.A. Johnston, Phys. Lett. B{\bf 286} (1992) 44;
Phys. Lett. B{\bf 291} (1992) 233.

\bibitem{CHEN-FERRENG-LANDAU-PRL-PRE-1992}
S. Chen, A.M. Ferrenberg, and D.P Landau, Phys. Rev. Lett. {\bf 69} (1992) 1213;
Phys. Rev. E {\bf 52} (1995)1377.

\bibitem{JANKE-VILLANOVA-NPBSUPPL-1995}
W. Janke and R. VIllanova, Nucl. Phys. B (Proc. Suppl.) {\bf 47} (1996) 641.

\bibitem{CARDY-JACOBSON-PRL-1997}
J. Cardy and J.L. Jacobsen, Phys. Rev. Lett. {\bf 79} (1997) 4063.

\bibitem{JANKE-WEIGEL-APPB-2003}
W. Janke and M. Weigel, Acta Phys. Polonica B {\bf 34} (2003) 4891.





\end{thebibliography}
\end{document}